\documentclass[]{aa}
\usepackage{graphicx}
\usepackage{txfonts}
\usepackage{natbib}
\bibpunct{(}{)}{;}{a}{}{,}

\def\jour#1#2#3#4{#1, {#2}, #3, #4}

\newcommand\lnp{{ Lecture Notes in Physics,\ }}
\def\prl{ Phys. Rev. Lett.}

\def\jgr{ J. Geophys. Res.}

\def\ssr{ Space Sci. Rev.}
\def\apj{ Astrophys. J.}
\def\apjl{ Astrophys. J. Lett.}
\def\apjs{ Astrophys. J. Supp.}
\def\APh{ A.Ph.}
\def\mnras{ M.N.R.A.S.}

\def\aap{ Astron. Astrophys.}

\begin{document}

\title{Monte Carlo simulations of a diffusive shock  with multiple scattering angular
distributions}

\author{Xin Wang\inst{1,2}\thanks{NAOC fellow: \email{wangxin@nao.cas.cn}.  This work was funded in part by
CAS-NSFC grant 10778605 and NSFC grant 10921303 and the National
Basic Research Program of the MOST (Grant No. 2011CB811401).} \and
Yihua Yan\inst{1}}

\institute{Key Laboratory of Solar Activities of National
Astronomical Observatories, Chinese Academy of Sciences,Beijing
100012, China; \and State Key Laboratory of Space Weather, Chinese
Academy of Sciences, Beijing 100080, China}

\abstract {} {We independently develop a simulation code following
the previous dynamical Monte Carlo simulation of the diffusive shock
acceleration under the isotropic scattering law during the
scattering process, and the same results are obtained.} {Since the
same results test the validity of the dynamical Monte Carlo method
for simulating a collisionless shock, we extend the simulation
toward including an anisotropic scattering law for further
developing this dynamical Monte Carlo simulation. Under this
extended anisotropic scattering law, a Gaussian distribution
function is used to describe the variation of scattering angles in
the particle's local frame.} {As a result, we obtain a series of
different shock structures and evolutions in terms of the standard
deviation values of the given Gaussian scattering angular
distributions. } {We find that the total energy spectral index
increases as the standard deviation value of the scattering angular
distribution increases, but the subshock's energy spectral index
decreases as the standard deviation value of the scattering angular
distribution increases.} 
\authorrunning{\it Xin Wang $\&$ Yihua Yan}
\titlerunning{\it Simulations of A Diffusive Shock } \maketitle

\section{Introduction}
\label{Introduction}

It is well known that the diffusive acceleration model has been
popular for more than five decades since \citet{fermi49} first
proposed that cosmic rays could be produced via diffusive processes.
Until now, diffusive shock acceleration (DSA) (i.e. first-order
Fermi acceleration) has been extensively applied to many physical
systems, such as shocks in the solar system, in our Galaxy, and
throughout the Universe \citep[][]{baring97,bell04}.

It is also well known that the nonlinear interactions in plasma
usually include such things as the turbulence of scattering wave
field, cosmic ray (CR) injection, and ``back reaction" by CR
pressure. These complex behaviors have held back comprehensive
understanding of the DSA and nonlinear DSA theory. Therefore, to
study the properties of the acceleration process and dynamical
behavior of the CR's ``back reaction" on the background flow,
choosing numerical simulation methods has been a primary and
essential problem. \citep{Ostrowski91,kang91, malkov97, ebj95,
ebj96, knerr96, bere94,slz10, bere00}. The main simulation methods
are introduced here, and a more detailed review can be seen in
\citet[][]{kang01}.

{\it Monte Carlo method}. In Monte Carlo simulations, the scattering
processes between individual particles with the collective
background flow are simulated around a one-dimensional parallel
shock. The particle scattering process is based on a prescribed
scattering law, and collecting moments based on the background
computational grid for scattering calculation is done by
particle-in-cell (PIC) techniques. \citet{knerr96} successfully
developed the dynamical Monte Carlo simulations for the Earth's bow
shock with important results in the maximum energetic particles
achieving greater than 1MeV accelerated by the shock. Before the
dynamical Monte Carlo simulation, \citet{emp90} had developed
stationary Monte Carlo simulations with the result that the cutoff
energy accelerated by the shock only reached 100 keV owing to the
limitation on the size of bow shock. \citet{baring95} also used the
stationary Monte Carlo method for simulating the oblique
interplanetary shocks, and the calculated results are a good fit to
the observed data. There are many works that use the Monte Carlo
method: \citet{ebj96}, \citet{ed02}, \citet{veb06,veb08}, and
others.

{\it Hybrid simulation}. The particles' motion equations are solved
explicitly based on the electromagnetic field of the background
plasma. Since the proton mass is about 2000 times that of the
electron mass, the total plasma is assumed to be a coupling of two
components: one component (e.g. electrons) is treated as a massless
fluid and the other component(e.g. protons,ions) is treated as
individual particles \citep{leroy82}. This method also employs the
PIC techniques based on computational grids. However, the limited
computational resources imply that the extensive calculation of the
electromagnetic field uses unrealistic parameters and are unable to
follow the shock for long enough \citep{giacalone93,giacalone09}.

{\it Two-fluid model}. The two-fluid model uses the
diffusive-convection equation, coupled with the gas dynamic
equations, to simulate the CR's acceleration as a gas component and
an accelerated particle component \citep{drury86, dorfi90, jones92}.
Since the CR energy density is solved instead of the particle's
distribution function in this model, the simulation results are
based on some assumptions, such as the CR's adiabatic index,
averaged diffusive coefficient, and injection rate. The averaged
diffusive coefficient needs to be inferred from the
diffusive-advection equation. The acceleration efficiency dependens
on the assumption of the injection rate. Under the reasonable a
priori models of these free parameters, a lot of simulations have
tested the acceleration efficiency and the shock structure, and both
agree with those derived from the diffusion-advection method. And
these semi-analytic solutions that have been extensively used can be
seen in recent works: \citet{mdv00}, \citet{bednarz01},
\citet{blasi1, blasi2}, \citet{ab05, ab06}, \citet{bac07},
\citet{cabv09}, \citet{cab10}.

{\it Kinetic simulation}. Within this full numerical simulation, the
diffusion-convection equation for the distribution function is
solved with a momentum-dependent diffusion coefficient  and a
suitable assumption of injection rate \citep{kang91, bere94,
bere00}. Unlike the two-fluid model,  the kinetic model should not
assume the CR's adiabatic index, in addition to using the
momentum-dependent diffusion coefficient instead of the averaged
diffusion coefficient. Berezhko and collaborators have studied
numerous DSA models for supernova remnants (SNRs) with the kinetic
model and conclude that about 20 \% of the total SN energy
transferred to the CR's population in the SNRs are more than the
results calculated from the two-fluid model. As for the energy
spectrum, the CR spectrum in this model shows a basic power-law
spectrum in the total energy range with a concave curve at some
energy range because of the precursor structure. More detailed new
studies of the kinetic model can be referred to in these papers
\citep{kjg02,kj07,abc08,lwsz09, za10}.

In an effort to follow and extend the  previous dynamical Monte
Carlo simulation \citep{knerr96}, we independently developed a
simulation code based on the Matlab platform using multiple
scattering laws. Our multiple scattering angular distributions
consist of three Gaussian distributions and one isotropic
distribution for the scattering angles during the scattering
process.  The aim of the isotropic scattering angular distribution
is to check the dynamical Monte Carlo method independently. Besides
this, we want to know how the Gaussian distributions affect the
scattering angular distribution function and the shock wave's
evolution and propagation; even more, we expect to find the
relationships between the multiple scattering law and the shock
compression ratio. To validate the multiple scattering angular
distributions, we followed the parallel-plane collisionless shock
and the particle's acceleration using the same parameters and data
as from Earth's bow shock, which was used in previous dynamical
Monte Carlo simulation.

In Section \ref{sec-propose}, we outline the motivations for
performing these four cases with four different scattering angular
distributions. In Section \ref{sec-method}, the specific simulation
techniques are described.  In Section \ref{sec-results}, we present
the shock simulation results and different cases with four
assumptions for scattering angle distributions. Section
\ref{sec-summary} includes a summary and the conclusions.
\section{The model}\label{sec-propose}

The dynamical Monte Carlo simulation has been developed by
\citet{knerr96} to study Earth's bow shocks. It gives  good results
for the higher than 1MeV cutoff in energy particles and the
power-law energy tail in the energy spectra. The dynamical Monte
Carlo simulation method uses the prescribed scattering law instead
of the complex electromagnetic field calculation like in the hybrid
model. In addition, the dynamical Monte Carlo simulation need not
assume the CR's injection rate and the associated diffusive
coefficient as do the two-fluid and kinetic models. For the above
reasons, we consider that developing a simulation code by following
the previous dynamical simulation is necessary. Although the
previous results successfully agree with observed data, the authors
mention that their results show that the total compression ratio of
the shock is more than 4, which should be less than the ratio of the
standard value for a nonrelativistic shock \citep{Pelletier01}. The
Rankine-Hugoniot (RH) jump conditions allow deriving the relation of
the compression ratio with the Mach number:
$r=(\gamma_{a}+1)/(\gamma_{a}-1+2/M^{2}) $, for a nonrelativistic
shock, the adiabatic index $\gamma_{a}$ = 5/3 , if the Mach number
$M \gg 1$, then the maximum compression ratio should be 4. To
validate these consistent results from the previous model and extend
this study to find what might be responsible for the shock
compression ratio, we extend the previous isotropic scattering
angular law by including an anisotropic scattering angular law. This
prescribed multiple scattering law consists of an isotropic
scattering angular distribution and an anisotropic scattering
angular distribution. The scattering angles consist of two variables
of pitch angle and azimuthal angle. Once a particle has a collision
with the massive scattering centers, its pitch angle becomes
$\theta'$=$\theta$+$\delta\theta$, and the azimuthal angle becomes
$\phi'$=$\phi$+$\delta\phi$, where $\delta\theta$ is the variation
in the pitch angle $\theta$, and $\delta\phi$ is the variation in
the azimuthal angle $\phi$. The pitch angles $\theta$ and $\theta'$
are both in the range $0\leq\theta,\theta'\leq \pi$, and azimuthal
angles $\phi$ and $\phi'$ are both in the range $0\leq\phi,\phi'\leq
2 \pi$ on the unit sphere. The variation in the pitch angle
$\delta\theta$ and azimuthal angle $\delta\phi$ are composed of the
scattering angle, and its anisotropic character is described by the
Gaussian function $f(\delta\theta,\delta\phi)$.

Under the multiple scattering angular distribution law, four cases
are calculated with three Gaussian distributions and one isotropic
random distribution for the scattering angles. Here, the sign
$\sigma$ is used to represent the standard deviation of the Gaussian
function, and the sign $\mu$ is used to represent the statistical
average or expected value of Gaussian function for the scattering
angles ($\delta\theta,\delta\phi$). We catalog the four cases as
follows.

 (1) Case A: the scattering angles ($\delta\theta,\delta\phi$)
 are distributed with a standard deviation $\sigma=\pi$/4 and an average
value $\mu=0$.

 (2) Case B: the scattering angles ($\delta\theta,\delta\phi$)
  are distributed with a standard deviation $\sigma=\pi$/2 and an average
value $\mu=0$.

 (3) Case C: the scattering angles ($\delta\theta,\delta\phi$)
  are distributed with a standard deviation $\sigma=\pi$ and an average
value $\mu=0$.

 (4) Case D: the scattering angles ($\delta\theta,\delta\phi$)
  are distributed with a standard deviation $\sigma=\infty$ and an average
value $\mu=0$, with $\delta\theta$ varying from $-\pi/2$ to $\pi/2$,
 and $\delta\phi$ varying from $-\pi$ to $\pi$ isotropically.

We performed four simulations according to the four different
assumptions of the scattering angular distributions algorithm. We
also assumed the scattering time (i.e., the mean time between two
scattering events) is the same constant in the four cases as in the
previous model. The idea that such a simple law can be used to
describe the entire scattering process was postulated by
\citet{eichler79}, based on the two-stream instability work done by
\citet{parker61}. Put simply, it is assumed that the turbulence
generated by both energetic particles streaming in front of the
shock  and by thermal particles produces nearly elastic scattering
for particles for all energies in diffusive shocks.

\section{Description of the method }\label{sec-method}

For simulating the total properties of shocks as they evolve from
formation to a final steady state as energy increases via Fermi
acceleration, we used the dynamical Monte Carlo model which employed
the PIC techniques. Because there is no assumption of the injection
rate or transparency function in PIC techniques, the shock-heated
downstream ions can freely scatter back across the subshock into the
upstream region without being thermalized, and the superthermal
particles are produced in the thermal background self-consistently.
In addition, unlike the hybrid simulation, there is no complicated
electromagnetic field calculation for individual particles, because
it is replaced by the prescribed scattering law \citep{ellison93}.
To reproduce these acceleration and scattering processes, a similar
simulation box and the same parameters (see Appendix
\ref{appen-schematic}) as the previous dynamical Monte Carlo method
are used in these new codes. As described in the previous simulation
(see Figure \ref{schematicfig}), the particles with an initial bulk
velocity $U_{0}$ and a Maxwellian thermal velocity $V_{L}$ in their
local frame are moving along a parallel magnetic field $B_{0}$ in a
one-dimensional box. To maintain a continuous flux-weighted flow
upstream, a new particle fluid with the same density  needs to be
injected into the simulation box from the left boundary. For a shock
initialization, the reflecting wall on the right boundary is used to
reflect the incoming particles, and forms a piston shock. The model
also includes the escape of energetic particles at the upstream free
escape boundary (FEB). The FEB phenomenologically models a finite
shock size or the lack of sufficient scattering far upstream to turn
particles around. Once ions cross the FEB, they are assumed to
decouple from the shock system, and are taken as the energy losses
\citep{je91}. The size of the foreshock region (the distance from
the shock front to the FEB) thus sets a limit on the maximum energy
a particle can obtain.

As shown in Figure \ref{schematicfig}, one particle's box frame
velocity $V$ is a total velocity, which is composed of the local
thermal velocity $V_{L}$ and the bulk fluid speed $U$ (i.e.
$V=V_{L}+U$, for upstream $U=U_{0}$, for downstream $U=0$). After
one particle arrives in the downstream region, its kinetic energy is
converted into random thermal energy by dissipation processes. With
the development of these many processes, the bulk fluid speed of
downstream flow becomes zero, and the length of downstream region is
extended dynamically.

As listed in Table \ref{parametertab}, all of the specific
parameters are used in our simulations, considering PIC techniques.
The total length ($X_{max}=300$) is divided into the number of grids
($n_{x}=600$) with a grid length ($\Delta x=1/2$). Initial grid
density of the particles ($n_{0}=650$) is set. The total time
($T_{max}=2400$) is divided into the number of time steps
($N_{t}=72000$) with an increment of time ($dt=1/30$). In summary,
these new codes consist of the following three substeps like the
previous simulation, except for the third substep employing the
extended multiple scattering laws.

 (i) Individual particles move. Particles with their
 velocities move along the one-dimensional $x$ axis:
\begin{equation}
X^{t}_{k}=X^{t-1}_{k}+(V_{x})^{t}_{k}dt, t \in [1,t_{max}],k\in
[1,k_{max}],\label{eq_step1a}\end{equation} where \begin{equation}
(V_{x})_{k}=(V_{Lx})_{k}+(U_{k})_{x},
\label{eq_step1b}\end{equation}

\begin{equation}
(U_{k})_{x}=\frac{1}{n_{k}}\displaystyle
\sum_{i=1}^{n_{k}}(V_{x})_{i}. \label{eq_step1c}\end{equation} here
$t_{max}$=72000, $k_{max}$=600, and ``k" represents the index of the
computational grid, $(U_{k})_{x}$ represents the bulk fluid speed of
the computational grid along to the $\hat{x}$ direction, and the
value of $U_{k}$ is obtained from substep (ii). Since we are
simulating a diffusive shock based on a one-dimensional parallel
magnetic field, the fluid quantities only vary in the $\hat{x}$
direction.

(ii) Mass collection. Summation of particle masses and velocities
are calculated  at the center of each computational grid:
\begin{equation}
P_{k}=\displaystyle \sum_{i=1}^{n_{k}} m_{p}(V_{x})_{i},
k=1,2,...k_{max}\label{eq_step2a}\end{equation}
\begin{equation}
U_{k}=\frac{1}{n_{k}}\displaystyle \sum_{i=1}^{n_{k}}(V)_{i},
k=1,2,...k_{max},\label{eq_step2b}\end{equation} where $n_{k}$ is
the number density of particles in the ``k'' grid, representing the
mass of the computational grid. Here, $P_{k}$ is the total momentum
of the protons in the ``k" grid, $m_{p}$ is the mass of an
individual proton, and $U_{k}$ is the average bulk fluid speed of
the grid (i.e. the velocity of the scattering center). The collected
grid-based mass and momentum densities will directly decide the
velocity of the scattering center $U_{k}$. The particle's total
velocity $V$ in the box frame is decided by Equation
(\ref{eq_step1b}). Once the value of $U_{k}$ becomes zero, the shock
front is decided by the position of the corresponding grid, and it
means that the shock is formed and the length of the downstream
region is extended dynamically. Similarly, if the value of $ U_{k}$
is between $U_{0}$ and zero, it means that the foreshock region or
precursor (i.e. between the FEB and the shock front) is formed by
the ``back pressure." The FEB and the shock front both dynamically
move away from the reflective wall with a shock velocity $v_{sh}$.

(iii)  Applying multiple scattering laws.  A certain fraction of the
particles are chosen to scatter the background scattering center
with their corresponding scattering angles according to the
prescribed scattering angular distributions. The average number of
scattering events occurring in an increment of time  $dt$ depends on
the scattering time scale $\tau$, and the scattering rate is
presented by
\begin{equation}
R_{s}=dt/\tau,\label{eq_step3a}\end{equation} where $R_{s}$ is the
probability of the scattering events occurring in an increment of
time. The candidates with their local velocities and scattering
angles scatter off the grid-based scattering centers. These
individual particles do not change their routes until they are
selected to scatter once again. So the particle's mean free path is
proportional to the local thermal velocities in the local frame with
\begin{equation}
\lambda \propto V_{L},\label{eq_step3c}\end{equation} for
simplicity, we take its formula as
\begin{equation}
\lambda = V_{L}\cdot \tau .\label{eq_step3b}\end{equation} For the
individual protons, the grid-based scattering center can be seen as
a sum of individual momenta. So these scattering processes can be
taken as the  elastic collisions. In an increment of time, once all
of the candidates complete these elastic collisions, the momentum of
the grid-based scattering center is changed.  In turn, the momentum
of the grid-based scattering center will affect the momenta of the
individual particles in their corresponding grid in the next
increment time. One complete time step consists of the above three
substeps. The total simulation temporally evolves forward by
repeating this time step sequence. To calculate the scattering
processes accurately and produce an exponential mean free path
distribution, the time step should be less than the scattering time
(i.e. $dt<\tau$).

The presented multiple scattering law simulations are developed on
the Matlab platform. Any one of the four cases can occupy the CPU
time for about seven weeks on a 3. 4GHZ (MF) CPU per core. To speed
up the running programs, the parallel algorithm should be used on a
high performance computer (HPC).


\section{Results \& discussions}\label{sec-results}
\begin{figure}\center
    \includegraphics[width=2.0in, angle=0]{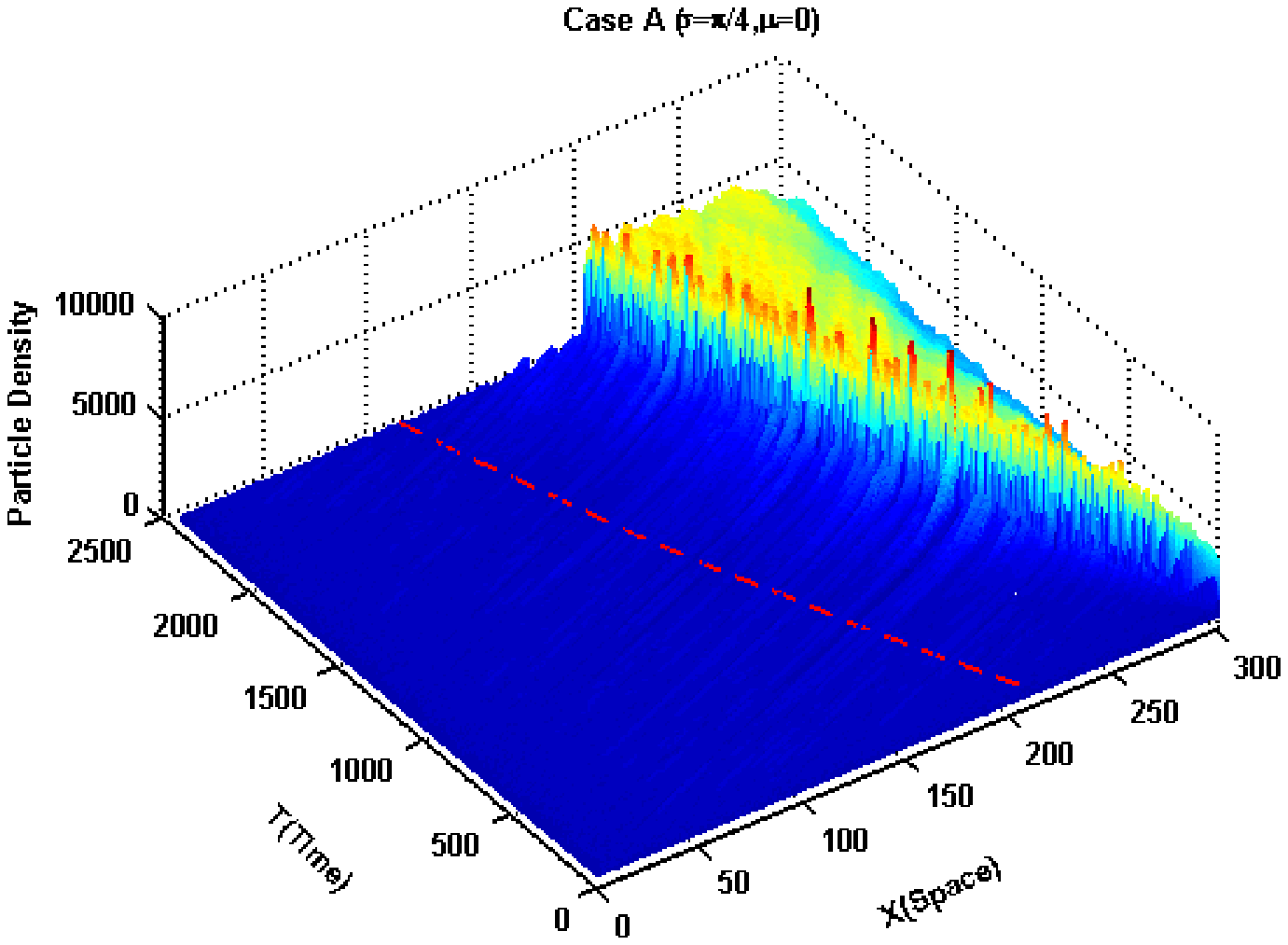}\\
    \includegraphics[width=2.0in, angle=0]{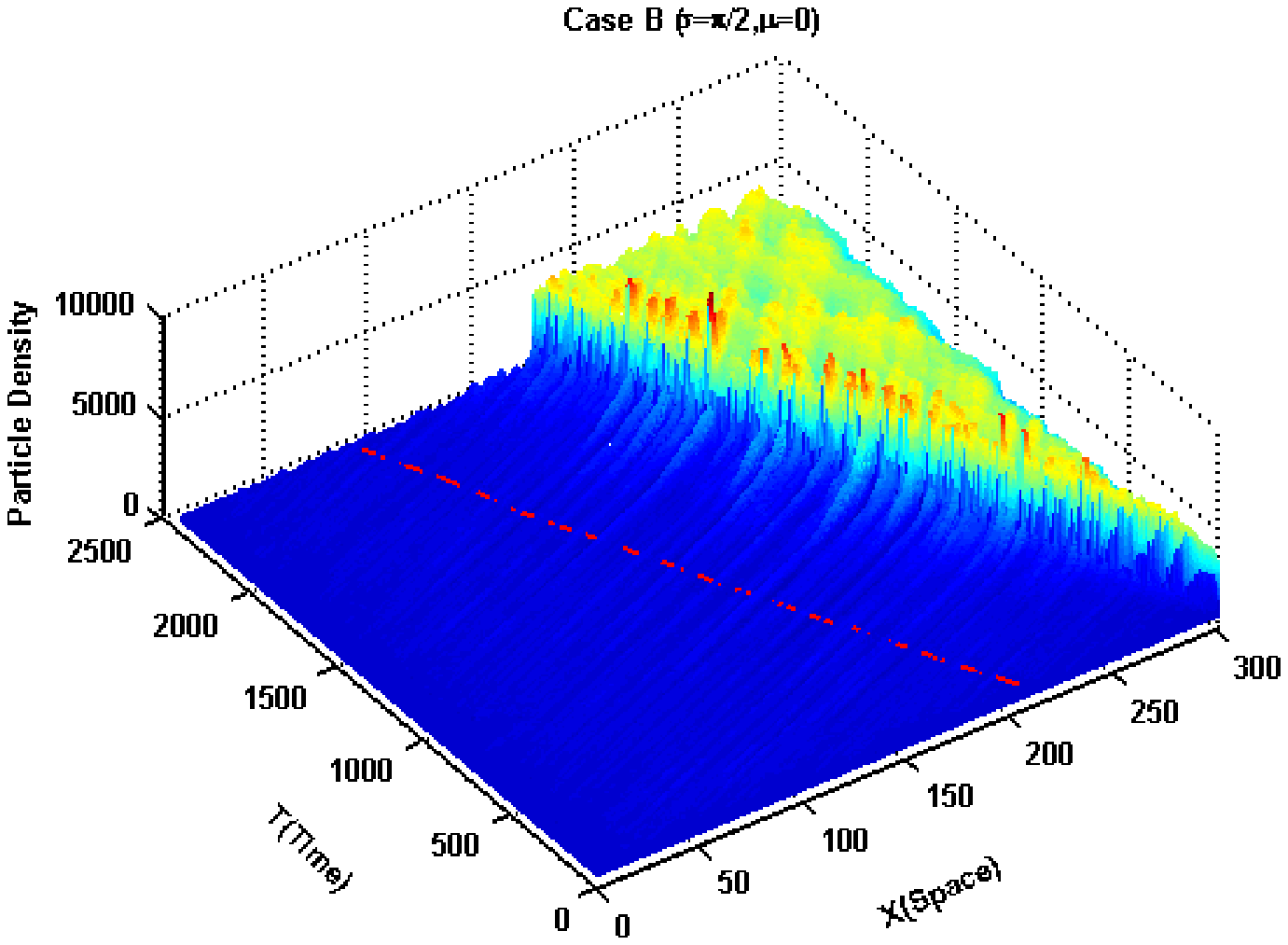}\\
    \includegraphics[width=2.0in, angle=0]{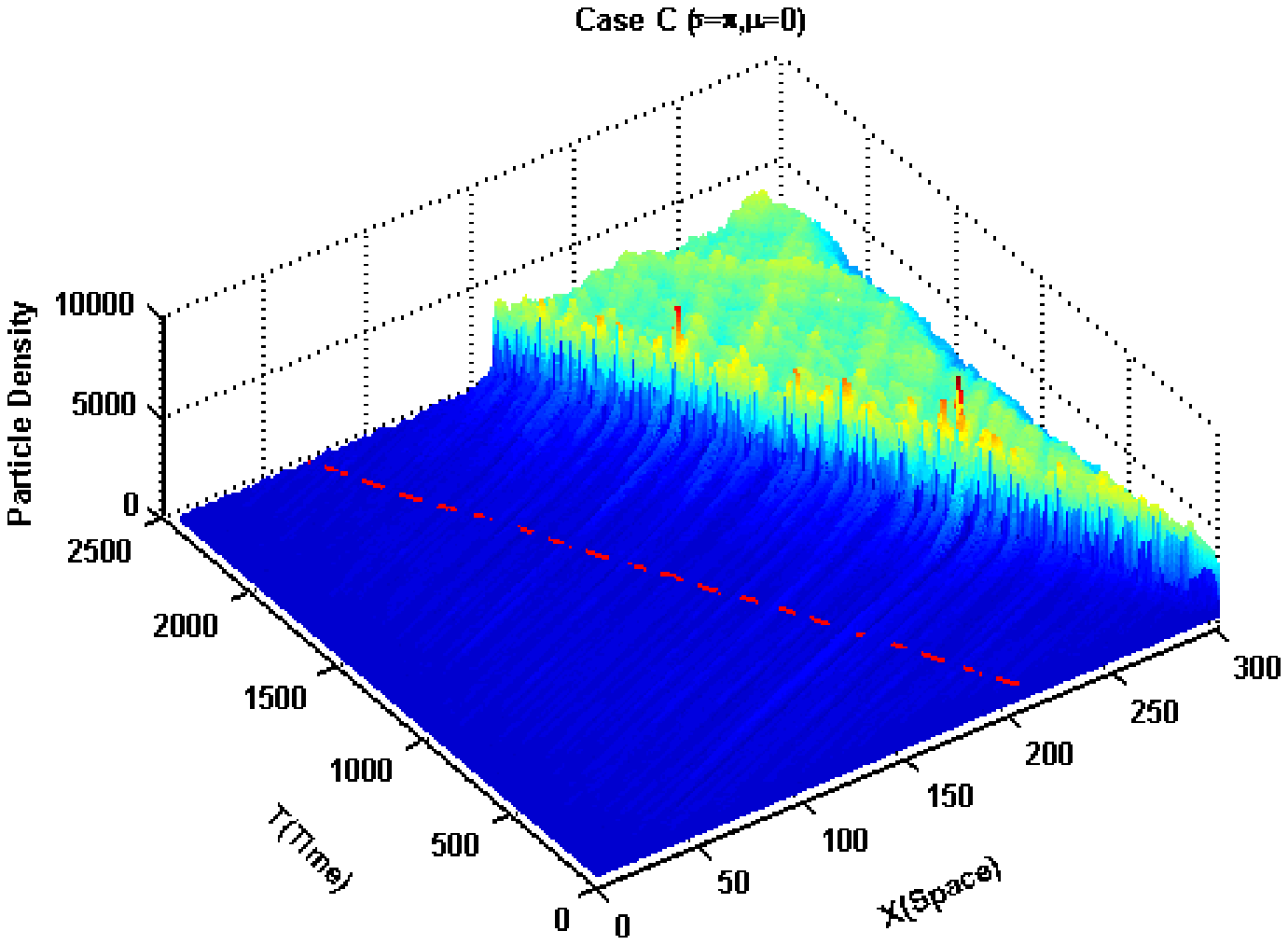}\\
    \includegraphics[width=2.0in, angle=0]{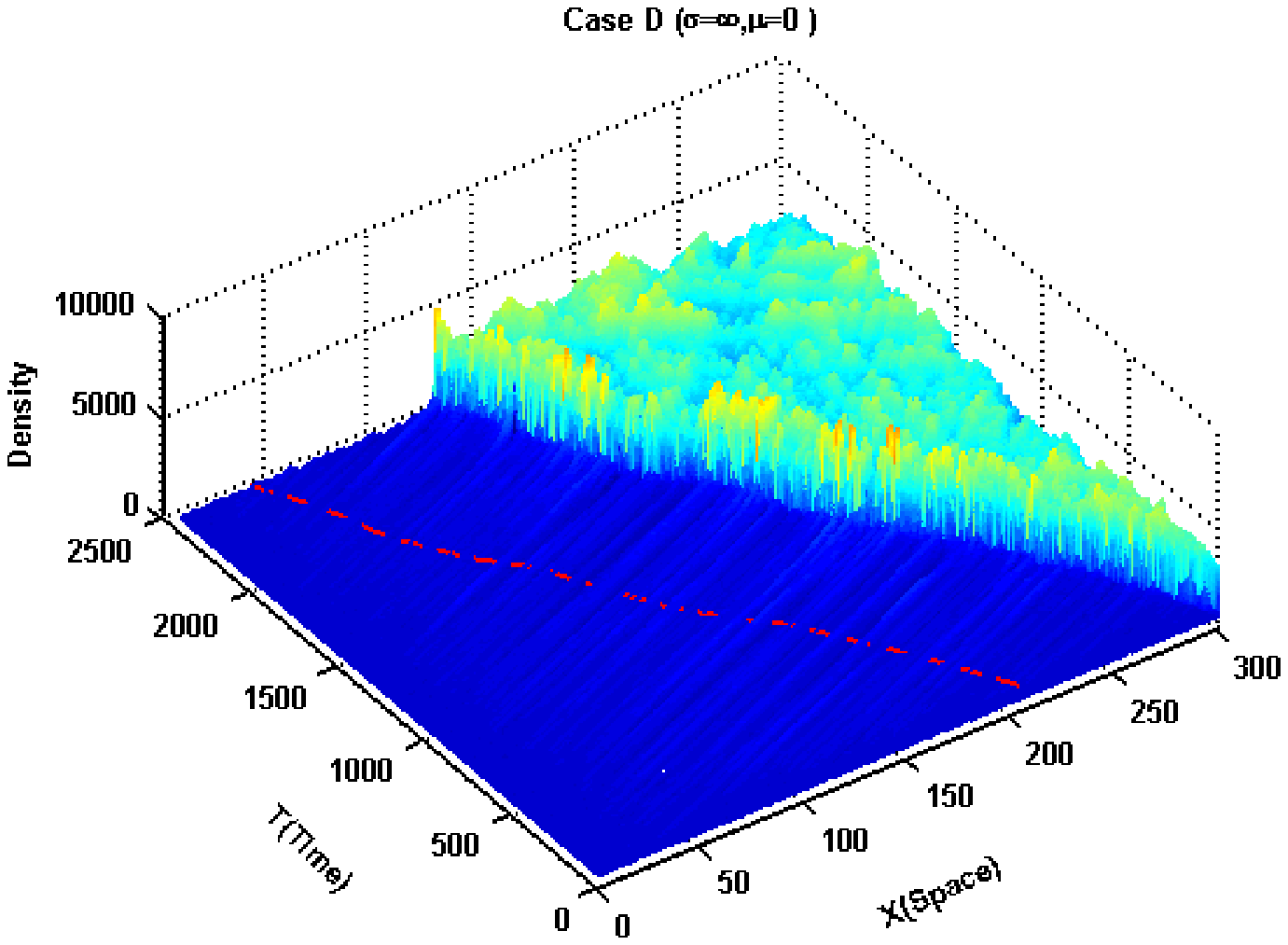}
\caption{Four cases of density profiles for the entire simulation
box vs the time. The dashed line represents the position of the FEB
in each plot.}\label{phasefig}
\end{figure}

We present all of the shock profiles for the shock simulations of
the four cases in Figure \ref{phasefig}, and we present all aspects
of simulation results including the density and velocity profiles,
compression ratios, analysis of the heating and acceleration, and
energy spectrum, as well as the correlations between the shock
compressions with the energy spectral index. For the convenience of
comparison and discussion, we list the specific calculated items in
Table \ref{restab}. Here, $\sigma$ is the standard deviation of the
scattering angular distribution function
$f(\delta\theta,\delta\phi)$, and $\mu$ is the average value of
scattering angles $(\delta\theta,\delta\phi)$. The subshock
compression ratios $r_{sub}$ are calculated from the velocity
profiles in the same shock frame reference. The compression ratios
$r_{u}$ and $r_{\rho}$  are calculated from velocity profiles and
density profiles, respectively. The total energy spectral index
$\Gamma_{tot} $ and the subshock's energy spectral index
$\Gamma_{sub} $ are calculated from compression ratios $r_{u}$ and
$r_{sub}$, respectively. The last two rows are shown as scaled
values.

\begin{table*}[]
\begin{center}
\caption{\label{restab} Results of calculation with an initial
energy of $E_{0}$=1.3105keV.}
\begin{tabular}{|c|c|c|c|c|c|c|c|c|c|c|c|c|c|c|c|c|c|c|c|c|c|}
  \hline   Items & Case A & Case B & Case C & Case D \\
  \hline  $f(\delta\theta,\delta\phi)$ & Gaussian & Gaussian & Gaussian & Isotropy\\
  \hline  $\sigma$ & $\pi/4 $& $\pi/2$ & $\pi$ & $\infty$ \\
  \hline  $\mu $& 0 & 0 & 0 & 0 \\
  \hline  $v_{sub}$ & 0. 1118 & 0. 1465 & 0. 1733 & 0. 2159 \\
  \hline  $v_{sh}$ & -0. 0433 & -0. 0535 & -0. 0617 & -0. 0733 \\
  \hline  $X_{sh}$ & 196 & 171. 5 & 152 & 124 \\
  \hline  $FEB$ & 106 & 81. 5 & 62 & 34 \\
  \hline  $\rho_{2} $& 5000 & 4225 & 3759 & 3303 \\
  \hline  $\rho_{1} $& 650 & 650 & 650 & 650 \\
  \hline  $r_{sub}$& 2. 5421 & 3. 0207 & 3. 3903 & 3. 9444 \\
  \hline  $r_{u} $& 7. 9231 & 6. 6031& 5. 8649 & 5. 0909 \\
  \hline  $r_{\rho} $& 7. 6936 & 6. 5001& 5. 7836 & 5. 0815 \\
  \hline  $\Gamma_{tot} $ & 0. 7167 & 0. 7677 & 0. 8083 & 0. 8667 \\
  \hline  $\Gamma_{sub} $ & 1. 4727 & 1. 2423 & 1. 1275 & 1. 0094 \\
  \hline  $V_{Lmax}$ & 10. 8251 & 16. 0001 & 17. 7824 & 20. 5286 \\
  \hline  $E_{cutoff}$ & 1. 10MeV & 2. 41MeV & 2. 98MeV & 4. 01MeV \\
  \hline  $|V_{sh}|$ & 58. 16km/s & 71. 87km/s & 82. 88km/s & 98. 46km/s \\
  \hline
\end{tabular}
\end{center}
\end{table*}

As shown in Figure \ref{phasefig}, the present isotropic Case D
largely appears similar to the results from previous dynamical
simulations by \citet{knerr96}.  In addition, all aspects of the
shock wave structure, density and velocity profiles, compression
ratio and  energy spectra present in isotropic Case D also give
similar results to the previous outcome. The specific results of the
present isotropic Case D are shown in the fifth column of Table
\ref{restab}: $v_{sh}$=-0.0733, $X_{sh}$=124, $FEB$=34,
$r_{sub}$=3.9444, $r_{u}$=5.0909, $r_{\rho}\sim r_{u}$ with a
difference of 0.18\%, $\Gamma_{sub}$=1.0094, $\Gamma_{tot}$=0.8667,
$|V_{sh}|$=98.46km/s, and $E_{cutoff}$=4.01MeV. As a comparison, the
corresponding results of the previous simulation are also given:
$v_{sh}$=-0.0720, $X_{sh}$=127.5, $FEB$=37.5, $r_{sub}$=3.20,
$r_{u}$=5.20, $r_{\rho}\sim r_{u}$ with a difference of 2.3\% ,
$\Gamma_{sub}$=1.1818, $\Gamma_{tot}$=0.8571, $|V_{sh}|$=96.9 km/s,
and $E_{cutoff}\sim$4.00MeV. Among comparable results, a slightly
larger differences in the values of the $r_{sub}$ and $\Gamma_{sub}$
between the two isotropic simulations would be distributed between
different sizes of the subshock region, decided differently. As seen
from the comparison of the results coming from two independent
simulation codes, the present simulation code successfully produced
good agreement in the results with those in previous dynamical Monte
Carlo simulation. Therefore, the present simulation code is based on
the Matlab platform  without using a supercomputer that can
independently validate the previous dynamical simulation method
using a completely different code for that supercomputer. Next, we
offer a series of discussions about the different cases considering
the specific aspects of the simulation results for diffusive shock.

\begin{figure}\center
  \includegraphics[width=2.0in]{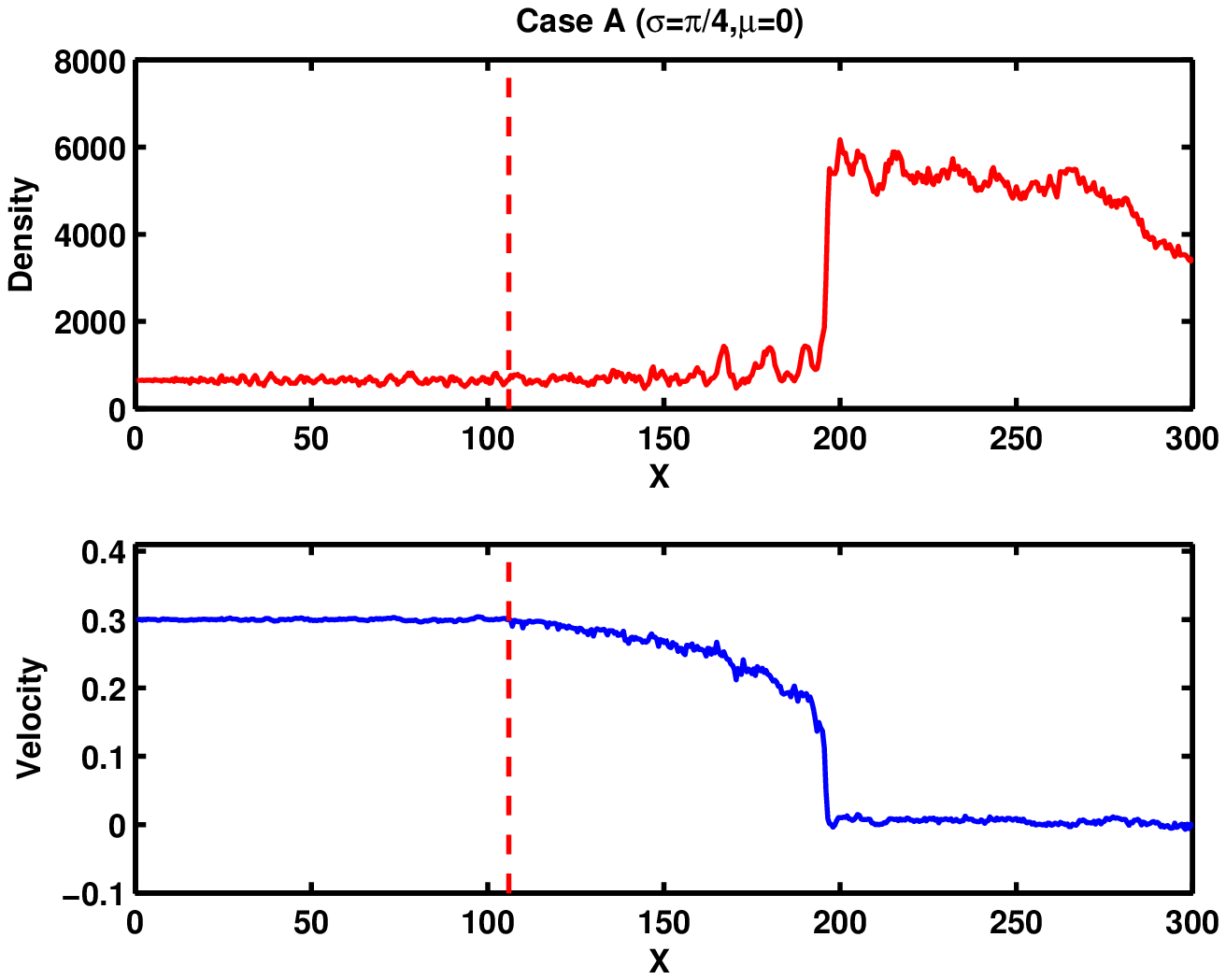}\\
  \includegraphics[width=2.0in]{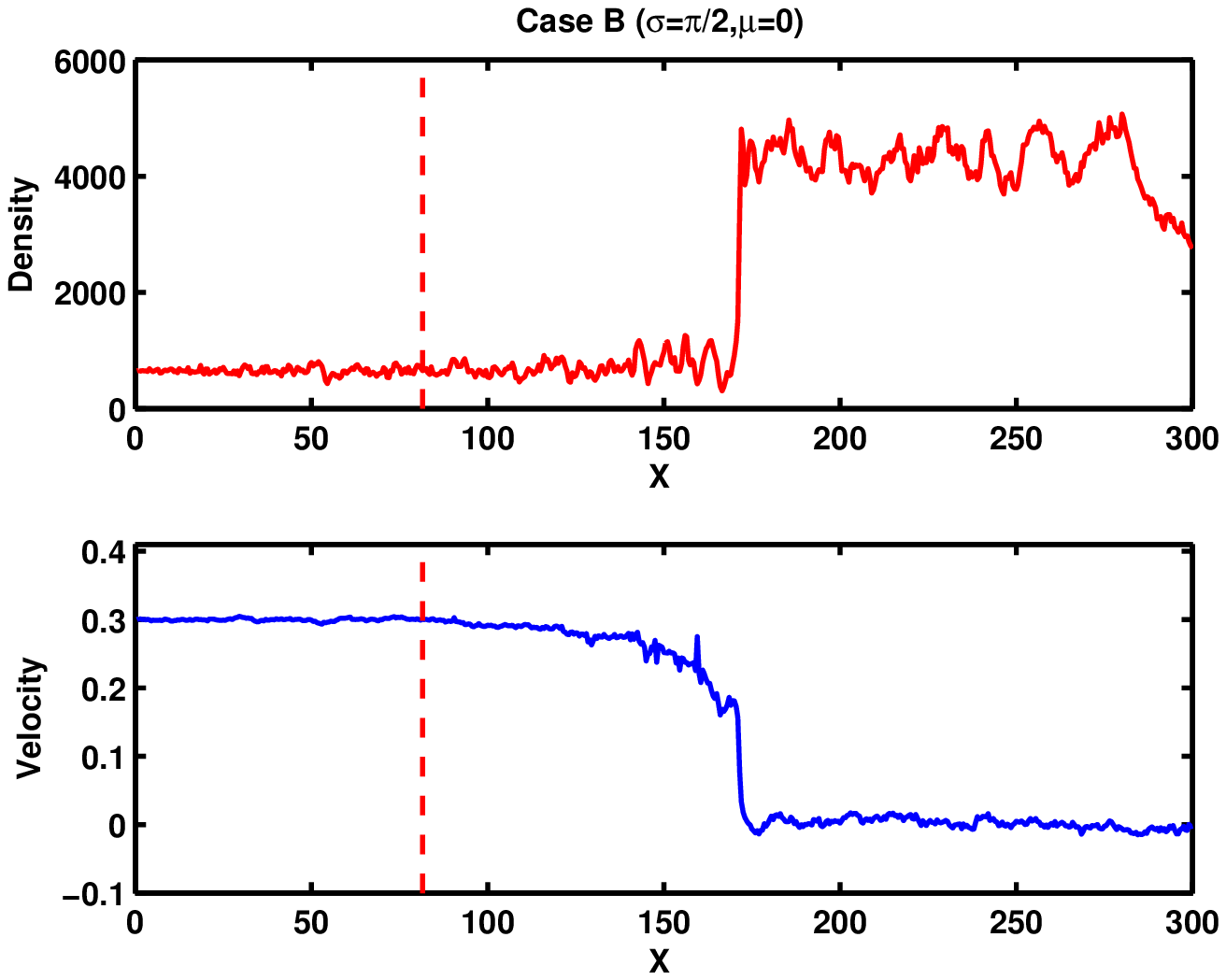}\\
  \includegraphics[width=2.0in]{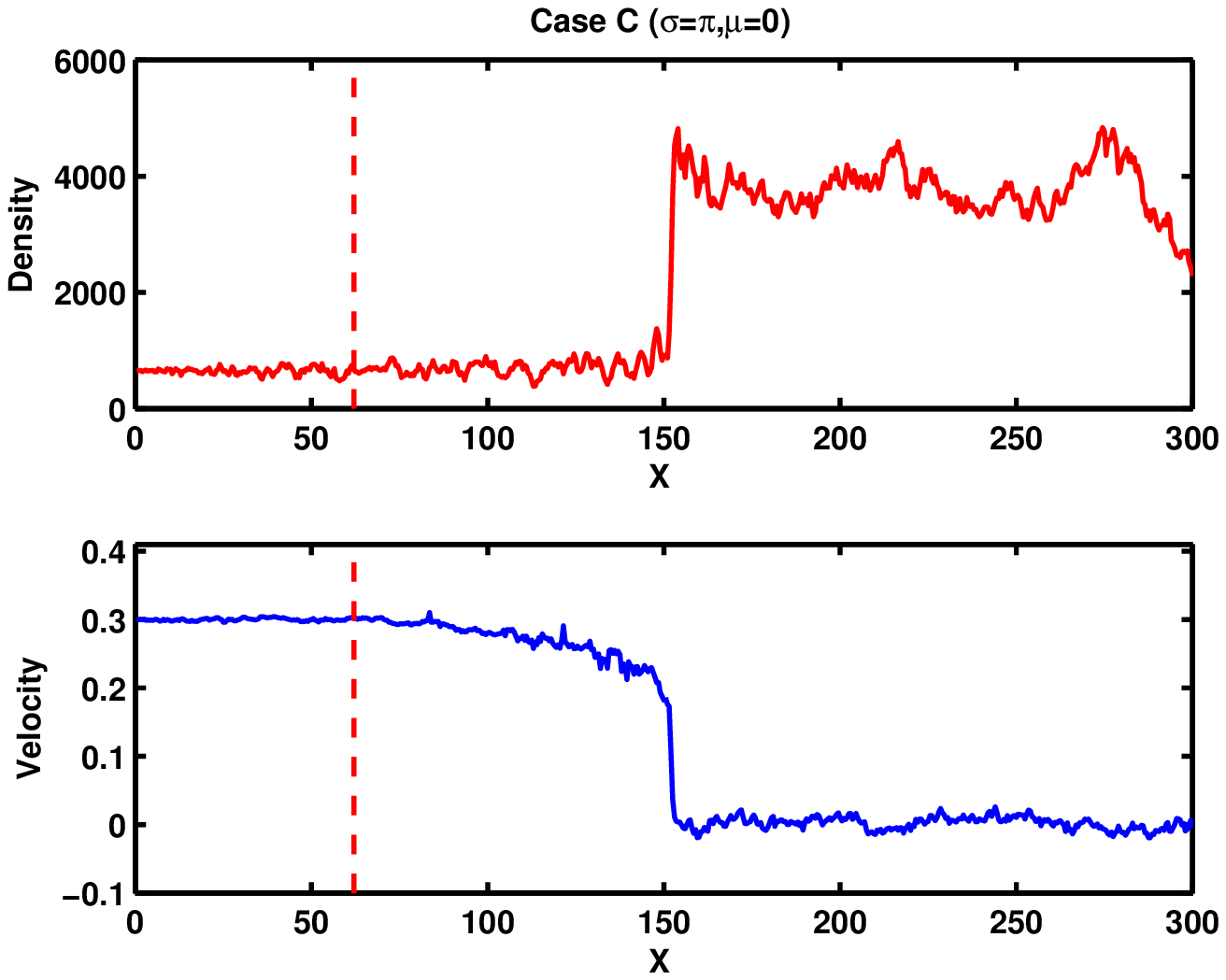}\\
  \includegraphics[width=2.0in]{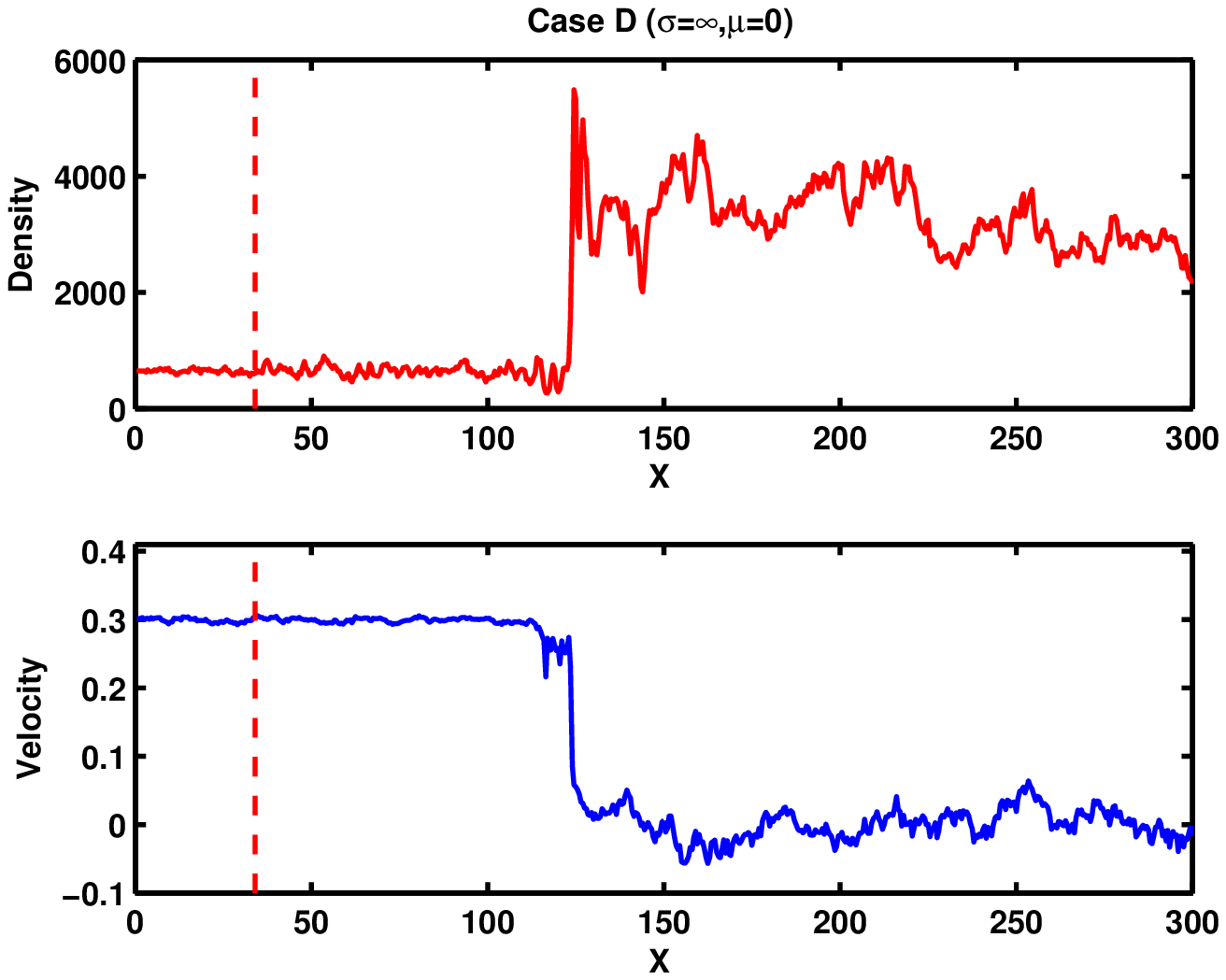}
\caption{The velocity and density profiles (in the box frame) vs the
position at the end of the simulation ($T_{max}$=2400)  in the four
cases. The vertical dashed line in each plot represents the position
of the FEB, from case A to case D.}\label{denfig}
\end{figure}
\subsection{Shock Profiles}
Figure \ref{phasefig} shows time sequences of the density profiles
of four cases. In each plot, a shock forms and moves away from the
reflective wall,  and the dashed line represents the FEB position
with the time in each case parallel to the shock front position. We
can see that both the shock position and the FEB position are moving
with a virtually constant velocity from the beginning of the
simulation to the end of the simulation (i.e. $T_{max}$=2400) in
each case. Simultaneously, as far as the positions of the FEB are
concerned, we can see that the FEB position at the end of the
simulation is significantly different in four different cases. As
for the average density fluctuation in the downstream region, there
are also apparent changes in different cases, and case A has the
slowest shock propagation speed among these four cases. Case D has
the lowest average density profile in the downstream region among
these four cases. Because from Cases A to D the only difference is
the prescribed scattering angular distribution, we conclude that
these differences of the results for shock propagation speed and
density profiles are decided by the standard deviation value
$\sigma$ of the scattering angular distribution.

Figure \ref{denfig} shows four cases of density and velocity
profiles at the end of the simulations.  From Cases A to D, the
position of the FEB approaches zero as the value of the standard
deviation $\sigma$ increases. The effects of the accelerated
particles are clearly seen in the upstream smoothing of the velocity
profiles in each case. In the simulations, when high-energy
particles cross the shock front and diffuse upstream, they
contribute negatively to the velocity profile. This reduces the
grid-based velocity in the zones upstream of the shock, which in
turn affects particles that are scattering in that region. In fact,
the accelerated particles slow and heat the incoming flow and smooth
the shock transition by their ``back reaction." As is obvious to see
from the velocity and density plots, different scattering angular
distributions  produce different effects on the shock wave
evolution. For the examples presented here, we consider that a
difference of approximately 40.93\% of the shock velocity is
contributed by the scattering angle distribution.

\subsection{Compression ratios}
Here, we compare  the compression ratios calculated from the
velocity profiles with those from the density relationships. First,
the value of the total compression ratio can simply be calculated
from the formula
\begin{equation}
r_{u}=u_{1}/u_{2}\label{eq_V_rho_a}
\end{equation}
where $u_{1}=u_{0}+|v_{sh}|$ , $ u_{2}=|v_{sh}|$, and $u_{1}(u_{2})$
is the upstream (downstream) velocity in the shock frame. The shock
velocity at the end of the simulation ($T_{max}$) can be derived
from the formula
\begin{equation}
 v_{sh}=(X_{max}-X_{sh})/T_{max}\label{eq_V_rho_a}
\end{equation} where $X_{max}$=300, $T_{max}$=2400, and $X_{sh}$ is the
position of the shock at the end of the simulation (see Table
\ref{restab}). The specific calculated results are shown in Table
\ref{restab}.

\begin{figure}\center
    \includegraphics[width=3.0in]{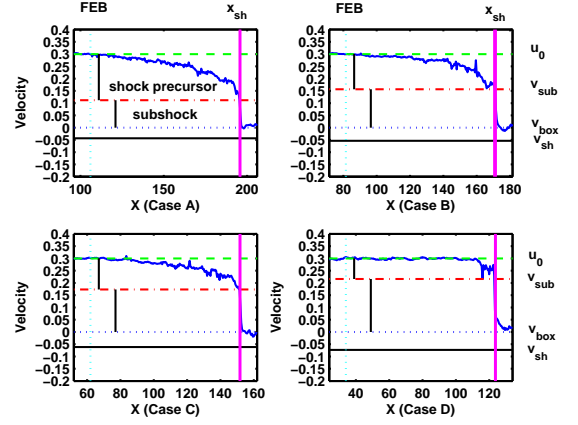}\\
  \caption{\small Velocity profiles in the shock region at the end of the
simulation ($T_{max}$=2400) in the four cases; the vertical solid
and dotted lines indicate the shock front and FEB in each plot,
respectively; the horizontal solid, dotted, dot-dashed and dashed
lines show the values of the shock velocity $v_{sh}$, velocity of
box frame $v_{box}$, subshock velocity $v_{sub}$ and initial bulk
velocity $U_{0}$, respectively. Two vertical bars in each plot
represent the two deflections of velocity, the upper bar represents
the part of the shock precursor and the lower one represents the
subshock.} \label{rsubfig}
\end{figure}

But in terms of the specific shock structure as seen in Figure
\ref{rsubfig}, an accurate subshock compression ratio calculation
should be more complicated. In any one of the cases in Figure
\ref{rsubfig} (plotted in the box reference frame), we show the
specific aspects of a shock modified by an energetic particle
population whose mean-free-path is an increasing function of
momentum.  The shock structure in each plot consists of three main
parts: precursor, subshock, and downstream.  The smooth precursor is
on the largest length scale between the FEB and near shock position
$X_{sh}$, where the fluid  speed gradually decreases from value
$U_{0}$ to $v_{sub}$. The size of this precursor is almost the
mean-free-path length of the maximum energy accelerated particles.
One of the smallest scales is the subshock region with a sharp
deflection of the fluid speed decreasing from $v_{sub}$ to
$v_{box}=0$. The downstream  region changes after the fluid speed
becomes $v_{box}=0$ by microphysical dissipation processes. The gas
subshock is just an ordinary discontinuous classical shock embedded
in the comparably larger scale energetic particle shock
\citep{bere99}. The value of $v_{sub}$ is determined by a sharp
deflection of smooth curves in velocity profiles near the shock
front, and the value of the subshock velocity increases from cases
A, B, and C to Case D (i.e.
$(v_{sub})_{A}<(v_{sub})_{B}<(v_{sub})_{C}<(v_{sub})_{D}$). All of
the velocity profiles are based on the box frame. That value of the
box frame's velocity is zero ($v_{box}=0$) in all cases. The
subshock compression ratio $r_{sub}$ is calculated from the formula
$r_{sub}= (v_{sub} +|v_{sh}|) /|v_{sh}|$ . For the sake of the
comparison of the values of $r_{sub}$ in different cases, the
subshock compression ratios are calculated in the same shock frame
reference, and the calculated results of $r_{sub}$ are shown in
Table \ref{restab}.

\begin{figure}\center
  \includegraphics[width=3.0in]{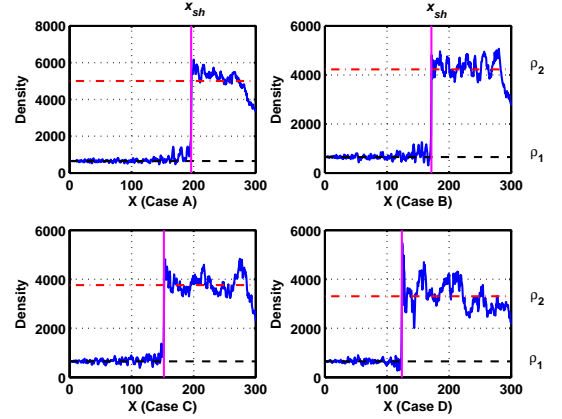}
\caption{Density profiles in an entire simulation box at the end of
the simulation ($T_{max}$=2400) in the four cases. The vertical
solid line is located in the position of the shock front, the upper
horizontal dot-dashed line represents the value of the downstream
density $\rho_{2}$,  and the lower horizontal dashed line indicates
the value of the upstream density $\rho_{1}$ in each plot. }
\label{nsubfig}
\end{figure}

We then have the calculations of the compression ratio from the
density relationships between the upstream and downstream flows:
\begin{equation}
r_{\rho}=\rho_{2}/\rho_{1}\label{eq_R_rho_a}
\end{equation}where $\rho_{1}=n_{0}$ is the upstream density, and
$\rho_{2}$ is the downstream density. This value is presented by
which is the average value of $n_{k}$ over the downstream region.
\begin{equation}
 \rho_{2}=\frac{1}{(k_{max}-k_{sh})}\displaystyle\sum_{k=k_{sh}}^{k_{max}}(n_{k})\label{eq_R_rho_b}
\end{equation} where $n_{k}$ is the
number density of particles in the ``k'' grid, $k_{sh}=x_{sh}/dx$ is
the grid index of the shock at the end of the simulation
($T_{max}=2400$), and $k_{max}=600$  is the grid index of the
$X_{max}$.

Figure \ref{nsubfig} shows the complete density plots of the four
cases at the end of the simulation. The value of the upstream
density $\rho_{1}$ is the same constant value, which is equal to the
initial density $n_{0}$ in each case.  The value of the downstream
density $\rho_{2}$ decreases from cases A, B, and C to Case D (i.e.
$(\rho_{2})_{A}>(\rho_{2})_{B}>(\rho_{2})_{C}>(\rho_{2})_{D}$).
Similarly, the detailed calculation results of the compression
ratios $r_{\rho}$ are listed in Table \ref{restab}. As listed in
Table \ref{restab},  the values of the subshock compression ratios,
$r_{sub}$=2.5421, $r_{sub}$=3.0207, $r_{sub}$=3.3903, and
$r_{sub}$=3.9444,  corresponding to cases A, B, C, and D,
respectively, are all lower than the standard value of $r=4$.
Unfortunately, the values of total compression ratio $r_{u} $ and
$r_{\rho} $ in each case are both higher than the standard value of
$r=4$. But \citet{knerr96} consider that, if energy is lost from the
system (e.g. by particles escaping via FEB), it is possible to
produce a shock with a total compression ratio that is higher than
the standard value predicted by the Rankine-Hugoniot (RH)
conditions. We have examined the mass and energy losses via the FEB
in each case. The results definitely show that the case with more
energy losses would produce a higher total compression ratio than
those in the case with less energy loss. Consequently, we consider
that the energy loss rates would be affected by the prescribed
scattering law. In any case, the energy losses are always an
important and interesting problem in the nonlinear diffusive shock
acceleration theory, so we will perform more precise research
focusing on these problems in later papers. In addition, although
the values of $r_{u} $ are correspondingly slightly higher than the
value of $r_{\rho} $ in each case, all these differences are less
than 3\%, and the specific difference in each case is
2.9\%,1.5\%,1.3\% and 0.18\% corresponding to cases A, B, C and D.
As seen from Figs. \ref{rsubfig} and \ref{nsubfig}, the value of the
total compression ratio $r_{tot}$, determined from the velocity
profiles, is more consistent with the density profiles in each case
(i.e. the total compression ratios $r_{tot}$ in all cases are
satisfied by the Rankine-Hugoniot (RH) conditions :
$u_{1}/u_{2}=\rho_{2}/\rho_{1}$). Therefore, it is not difficult for
us to conclude that the total compression ratios $r_{u} $ and
$r_{\rho} $  decrease as the value of the standard deviation
$\sigma$  of the scattering angular distribution increases, but the
subshock compression ratio $r_{sub}$ increases as the value of the
standard deviation $\sigma$ of the scattering angular distribution
increases.

\subsection{Heating \& acceleration}
\begin{figure}\center
 \includegraphics[width=3.0in]{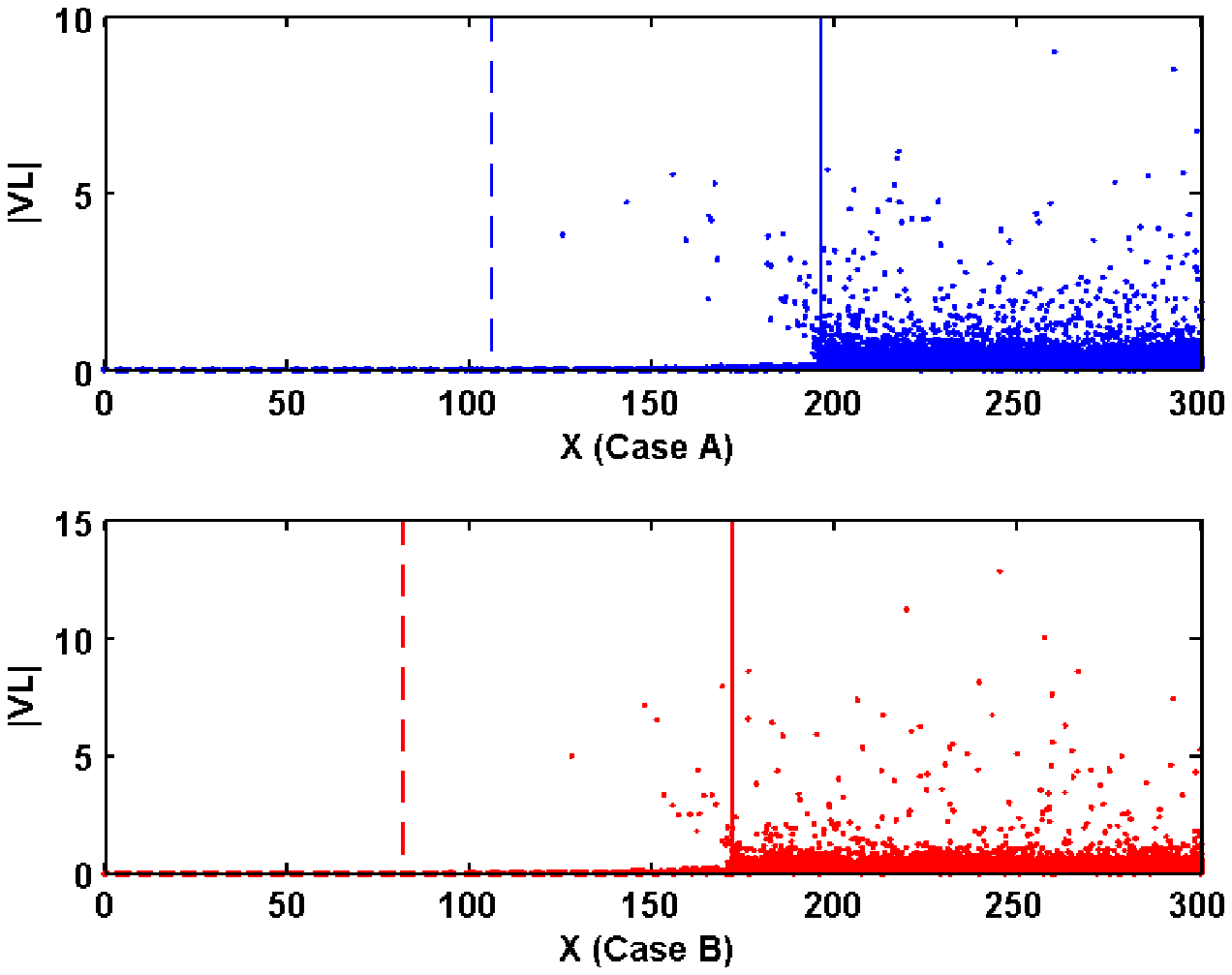}\\
  \includegraphics[width=3.0in]{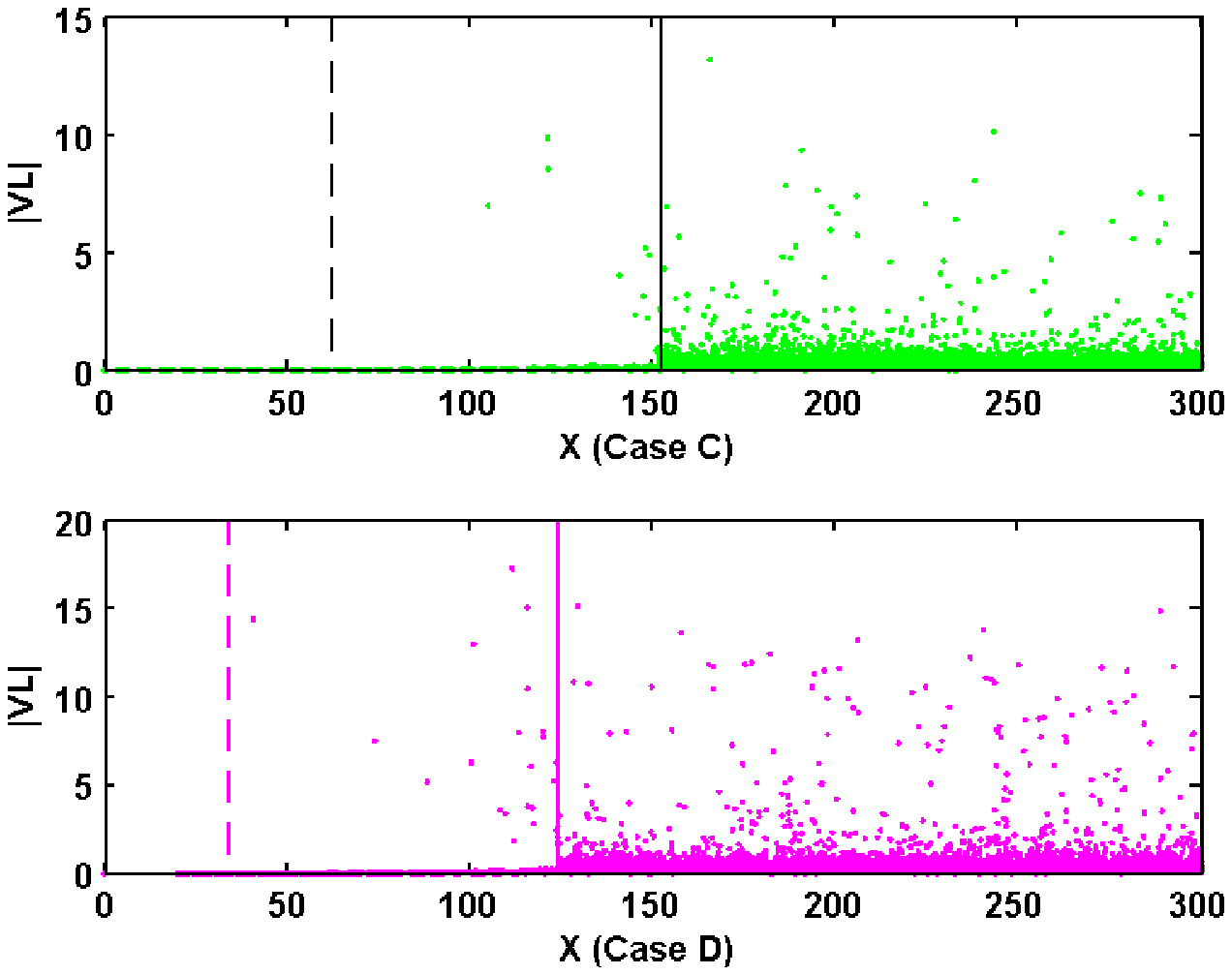}
  \caption{The scatter plots of the particle's thermal velocities in the local frame
  vs its position at the end of the simulations ($T_{max}$=2400),  and the vertical dashed line  and
   solid line in each plot indicate the approximate position
of the FEB and shock front, respectively. Only the ratio of 1/12 of
the total number of particles are plotted.}\label{scatterfig}
\end{figure}
Here we contrast between two important aspects of the heating and
acceleration processes in the diffusive shock acceleration. Figure
\ref{scatterfig} shows the particles' scatter plots in four cases at
the end of the simulations ($T_{max}$=2400). In each case, particles
with local velocity scatter into the simulation box's position. A
large amount of particles that do not get injected into the Fermi
acceleration mechanism and  that have lower thermal velocities stay
in the downstream region, and a few particles with higher energy via
multiple shock encounters can move far away from the shock front and
even escape from the FEB. From Cases A to D, more and more particles
are injected into the Fermi acceleration mechanism, and they gain
greater and greater maximum energy. Obviously, the maximum thermal
velocity in Case D would be several times that of the ones in Case
A. We supposed that this difference is mainly contributed by the
scattering angular distribution function
$f(\delta\theta,\delta\phi)$. In short, the majority of particles,
which flow toward the shock, cross the shock only once, and  (after
scattering) remain fairly stationary in the downstream region, which
would consist of the ``heated" elements, and a few high-energy
particles represent the ``power-law" part of the simulated particles
flows. Actually, the ``back pressure" from the accelerated particles
via their back reaction reduces the incoming fluid speed, leading to
a smoothed precursor heating. Therefore, an anisotropic scattering
angular distribution in the shocks dominate the ``gas heating"
process in the simulated plasma, and isotropic scattering angular
distribution in the shocks play an important role in nonlinear
acceleration of the energetic particles via the Fermi mechanism, and
they dominate the ``precursor heating".
\subsection{Energy spectra}
\begin{figure}\center
    \includegraphics[width=3.0in]{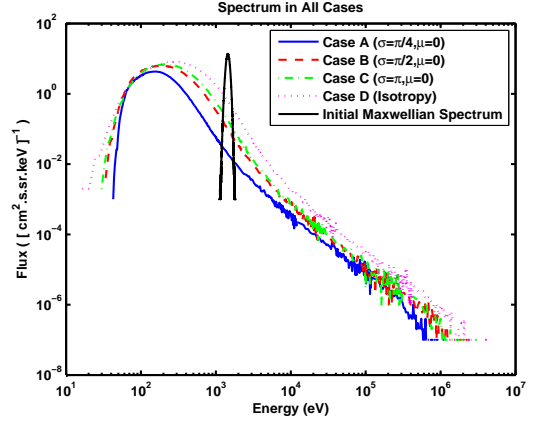}
\caption{The final energy spectrum in the four cases and the initial
energy spectrum plots. The thick solid line with a narrow peak at
$E_{0} = $1.3105keV represents the initial Maxwell energy
distribution. The solid, dashed, dot-dashed, and dotted extended
curves indicate the simulated particles' energy spectral
distribution, averaged over the entire downstream region, at the end
of the simulations ($T_{max}$=2400), corresponding to Cases A, B, C,
and D, respectively. Most particles cross the shock only once,
producing the large broad peak centered at E$_{A}\sim$ 0.05keV,
E$_{B}\sim $0. 1keV, E$_{C}\sim$ 0.15keV, and E$_{D}\sim $0.20keV in
Cases A, B, C, and D, respectively. However, some particles gain
enough energy via the Fermi acceleration mechanism to produce the
``power-law" tail in the energy spectrum with the cutoff at $E_{A}
$=1.10 MeV, $E_{B}$=2.41 MeV, $E_{C}$=2.98 MeV, and $E_{D}$=4.01 MeV
corresponding to Cases A, B, C, and D. All these energy spectra are
plotted in the same shock frame. }\label{spectrumfig}
\end{figure}
Spectra calculated in the shock frame from the initial and final
particle  (proton)  energy distributions in all cases are shown in
Figure \ref{spectrumfig}.  The energy units in this plot are derived
from the scaling parameters presented in Table \ref{parametertab}
(see Appendix \ref{appen-schematic}). Initially, all particles move
toward the wall with a certain thermal spread in energy. A narrow
peak at E=1.3105keV represents the initial Maxwell energy
distribution. The four extended curves indicate the simulated
particle energy spectral distribution, averaged over the entire
downstream region, at the end of the simulations, corresponding to
the four cases, respectively. The majority of the particles cross
the shock only once, producing an expanded energy spectrum with a
central peak at E$_{A}\sim$ 0.05keV, E$_{B}\sim $0.1keV, E$_{C}\sim$
0.15keV, and E$_{D}\sim $0.20keV in Cases A, B, C, and D,
respectively. However, as is shown in Figure \ref{spectrumfig}, the
minority of the particles gain enough energy via the Fermi
acceleration mechanism to produce the ``power-law" tail in the
energy spectrum with the cutoff at $E_{A} $=1.10 MeV, $E_{B}$=2.41
MeV, $E_{C}$=2.98 MeV and $E_{D}$=4.01 MeV corresponding to Cases A,
B, C and D, respectively. For more details about the calculated
results, see Table \ref{restab}. It is evident from Figure
\ref{spectrumfig} that the values of the central peak of the
extended energy spectra in the four cases are far from the initial
energy peak in their respective order. This means the values of the
central peak in each case increase as the value of the standard
deviation of the scattering angular distribution increases, and each
extended curve shows a harder power-law slope in its high-energy
tail as the expand energy range increases, respectively. Therefore,
we can see that the case of applying an anisotropic scattering
angular distribution function will produce a slightly softer energy
spectrum, and the case of applying an isotropic scattering angular
distribution will produce a slightly harder energy spectrum.

\subsection{Spectral index \& compression ratios}
\begin{figure}[h]\center
    \includegraphics[width=2.5in]{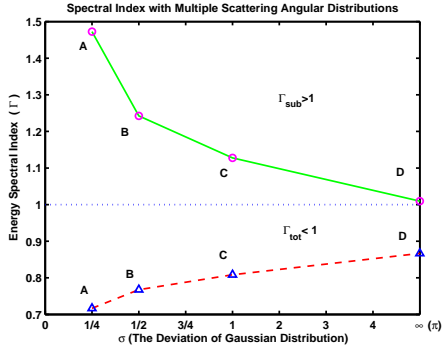}
\caption{The correlation of the deviation value of the Gaussian
distribution vs the energy spectral index. The triangles represent
the  total energy spectral index in each case. The circles indicate
the subshock's energy spectral index in each case. From Cases A to
D, all of the values of the subshock's energy spectral index
$\Gamma_{sub}> 1$ show a slightly harder power-law slope in the
respective order.  From Cases A to D, all of the values of the total
energy spectral index $\Gamma_{tot}< 1$  show a slightly decreasing
deviation from the power-law slope in the respective energy
spectrum. }\label{indexfig}
\end{figure}

Usually, we could predict the power-law energy spectral index from
diffusive shock acceleration theory:
\begin{equation}
dJ/dE\propto E^{-\Gamma}\label{eq_index_a}
\end{equation}
where $dJ/dE$ is the energy flux and $\Gamma $ is the energy
spectral index, and
\begin{equation}
\Gamma = (r+2) /(2\times (r-1)).\label{eq_index_b}
\end{equation}
According to Equation \ref{eq_index_b}, we substituted the values of
the compression ratio $r$ with two group values of $r_{tot}$ and
$r_{sub}$ obtained in each case. Then, the two group energy spectral
indices $\Gamma_{tot}$ and $\Gamma_{sub}$ in  each case are
calculated. Two groups' spectral index values are listed in Table
\ref{restab} as $\Gamma_{A}$= 0.7167, $\Gamma_{B}$ =0.7677,
$\Gamma_{C}$ =0.8083, and $\Gamma_{D}$=0.8667  in the total group
$\Gamma_{tot} $, and $\Gamma_{A}$= 1.4727, $\Gamma_{B}$ =1.2423,
$\Gamma_{C}$ =1.1275, and $\Gamma_{D}$=1.0094 in subshock group
$\Gamma_{sub} $, corresponding to the cases A, B, C, and D,
respectively. As shown in Figure \ref{indexfig}, from Cases A to D,
all of the values of the subshock's energy spectral index
$\Gamma_{sub}> 1$ and show that a slightly harder power-law slope in
the respective order.  From Cases A to D, all of the values of the
total energy spectral index $\Gamma_{tot}< 1$ show a slightly
decreasingly deviation from the power-law slope in the respective
energy spectrum.

\section{Summary and conclusions}\label{sec-summary}
We followed the previous dynamical Monte Carlo simulation by using a
new code based on the Matlab platform independently, and presented
the same results as the outcome from the previous simulation. In
addition, we successfully extended the simulation include the
multiple scattering angular distributions using these new codes to
study the diffusive shock acceleration mechanism further.

In conclusion, the comparison of the calculated results come from
different extensive cases, we find that the total energy spectral
index increases as the standard deviation value of the scattering
angular distribution increases, but the subshock's energy spectral
index decreases as the standard deviation value of the scattering
angular distribution increases. In these multiple scattering angular
distribution simulation cases, the prescribed scattering law
dominates the shock structure and plays an important role in
balancing whether the particles have more heating or more
acceleration. In other words, the cases with anisotropic scattering
distribution give the overall velocity-deflection precursor sizes,
which are larger than the isotropic case, and give a relatively
greater ``heating" effect or less ``acceleration" effect on
background flows than does the isotropic case.

As a result, the shock compression ratio  and the energy spectral
index are both modified naturally  by the prescribed scattering law.
Specifically, the cases of applying  an anisotropic scattering
distribution function will produce a slightly softer subshock's
energy spectral index, and the case of applying an isotropic
scattering angular distribution will produce a slightly harder
subshock's energy spectral index. Simultaneously, from the isotropic
case to the anisotropic case, the total energy spectrum shows an
increasing deviation from the ``power-law" distribution.

In addition, although we find no case producing the total
compression ratio which should be less than the standard value 4
according to the Rankine-Hugoniot (RH) jump conditions, the fact is
clear that the prescribed scattering angular distribution function
would have an effect on the total compression ratio. If there is a
suitably prescribed scattering law that leads to  much less energy
loss, it is possible to constrain the total compression ratio to be
less than 4.

\acknowledgements{The authors would like to thank Doctors G. Li,
Hongbo Hu, Siming Liu, Xueshang Feng, and Gang Qin for many useful
and interesting discussions concerning this work. In addition, we
also appreciate Profs. Qijun Fu and Shujuan Wang, as well as other
members of the solar radio group at NAOC.}


\begin{thebibliography}{99}\small
\bibitem[\protect\citeauthoryear{Amenomori et al.}{2008}]{abc08}
Amenomori, M.; Bi, X. J.; Chen, D. \& coworkers, \ 2008  \apj, 678,
1165.
 \bibitem[\protect\citeauthoryear{Amato \& Blasi}{2005}]{ab05}
    Amato, E. \& Blasi, P., 2005, \mnras Lett., 364, 76
\bibitem[\protect\citeauthoryear{Amato \& Blasi}{2006}]{ab06}
    Amato, E. \& Blasi, P., 2006, \mnras, 371, 1251
\bibitem[\protect\citeauthoryear{Bednarz \& Ostrowski}{2001}]{bednarz01}
Bednarz, J. \& Ostrowski, M.,  \ 2001, \mnras, 310, L13.
\bibitem[\protect\citeauthoryear{Baring, Ellison \& Jones }{1995}]{baring95}
 Baring, M. G., Ellison, D. C. \& Jones, F. C., \ 1995, Adv. Space Res., 15, 397
\bibitem[\protect\citeauthoryear{Baring}{1997}]{baring97}
Baring, M. G., \ 1997, Very High Energy Phenomena in the Universe;
Morion Workshop, Edited by Y. Giraud-Heraud and J. Tran Thanh Van,
1997., p.97
\bibitem[\protect\citeauthoryear{Bell}{2004}]{bell04}
Bell, A.R.,  \ 2004,  \mnras,  353, 550
\bibitem[\protect\citeauthoryear{Berezhko et al.}{1994}]{bere94}
Berezhko E.G., Yelshin V.K. \& Ksenofontov L.T. 1994, Astropart.
Phys., 2, 215
\bibitem[\protect\citeauthoryear{Berezhko \& Ellison }{1999}]{bere99}
 Berezhko E.G. \& Ellison, D.~C. 1999, \apj, 526, 385
\bibitem[\protect\citeauthoryear{Berezhko \& V\"olk}{2000}]{bere00}
 Berezhko E.G. \& V\"olk H.J. 2000, \aap, 357, 283
 \bibitem[\protect\citeauthoryear{Blasi}{2002}]{blasi1}
    Blasi, P., 2002, \APh, 16, 429
\bibitem[\protect\citeauthoryear{Blasi}{2004}]{blasi2}
    Blasi, P., 2004, \APh, 21, 45
\bibitem[\protect\citeauthoryear{Blasi, Amato \& Caprioli}{2007}]{bac07}
    Blasi, P., Amato, E. \& Caprioli, D., 2007, \mnras, 375, 1471
 \bibitem[\protect\citeauthoryear{Caprioli et al.}{2009a}]{cabv09}
Caprioli, D., Blasi, P., Amato, E. \& Vietri, M., 2009a, \mnras,
395, 895
\bibitem[\protect\citeauthoryear{Caprioli, Amato \& Blasi}{2010a}]{cab10}
Caprioli, D., Amato, E. \& Blasi, P., 2010a, \APh, 33, 160
\bibitem[\protect\citeauthoryear{Drury \& Falle}{1986}]{drury86}
Drury, L.~O'C. \& Falle, S.~A.~E.~G. 1986, \mnras, 223, 353
\bibitem[\protect\citeauthoryear{Dorfi}{1990}]{dorfi90}
Dorfi, E.A., 1990, \aap, 234, 419
\bibitem[\protect\citeauthoryear{Ellison, M\"obius \& Paschmann}{1990}]{emp90}
Ellison, D.~C., M\"obius, E. \& Paschmann, G.
 \jour{1990}{\apj}{352}{376}
\bibitem[\protect\citeauthoryear{Ellison  et.al.}{1993}]{ellison93}
Ellison, D.~C., Giacalone, J. , Burgess, D., \& Schwartz, S.~J.,
1993, \jgr, 98, 21058
\bibitem[\protect\citeauthoryear{Ellison, Baring \& Jones}{1995}]{ebj95}
Ellison, D.~C., Baring, M.~G. \& Jones, F.~C., 1995, \apj, 453, 873
\bibitem[\protect\citeauthoryear{Ellison, Baring \& Jones}{1996}]{ebj96}
Ellison, D.~C., Baring, M.~G. \& Jones, F.~C., 1996, \apj, 473, 1029
\bibitem[\protect\citeauthoryear{Ellison \&     Double}{2002}]{ed02}
Ellison, D.~C. \& Double, G.~P., 2002, \APh, 18, 213
\bibitem[\protect\citeauthoryear{Eichler}{1979}]{eichler79}
Eichler, ~D., \ 1979, \apj, 229, 419
\bibitem[\protect\citeauthoryear{Fermi}{1949}]{fermi49}
 Fermi, E., \ 1949, \prl 75, 1169
\bibitem[\protect\citeauthoryear{Giacalone et al.}{1993}]{giacalone93}
Giacalone, ~J., Burgess, ~D, Schwartz, S.~J. \& Ellison, D.~C., \
1993, \apj, 402, 550
\bibitem[\protect\citeauthoryear{Giacalone \& Jokipii}{2009}]{giacalone09}
Giacalone, ~J. \& Jokipii, J. R.  \ 2009, \apj, 701, 1865.
\bibitem[\protect\citeauthoryear{Pelletier}{2001}]{Pelletier01}
Pelletier, G. \ 2001, \lnp, 576, 58
\bibitem[\protect\citeauthoryear{Jones \& Ellison}{1991}]{je91}
Jones, F.~C. \& Ellison, D.~C. \jour{1991}{\ssr}{58}{259}.
\bibitem[\protect\citeauthoryear{Jones \& Kang}{1992}]{jones92}
Jones, T. W. \& Kang, H. 1992, \apj, 396, 575
\bibitem[\protect\citeauthoryear{Kang \& Jones}{1991}]{kang91}
 Kang H. \& Jones T.W. 1991, \mnras, 249, 439
\bibitem[\protect\citeauthoryear{Kang}{2001}]{kang01}
Kang, H., \ 2001, \apj, 550, 737
\bibitem[\protect\citeauthoryear{Kang, Jones \& Gieseler}{2002}]{kjg02}
    Kang, H., Jones, T.~W. \& Gieseler, U.D.J., 2002, \apj 579, 337
\bibitem[\protect\citeauthoryear{Kang \& Jones}{2007}]{kj07}
    Kang, H., Jones, T.~W., 2007, \APh, 28, 232
\bibitem[\protect\citeauthoryear{Knerr, Jokipii \& Ellison}{1996}]{knerr96}
Knerr, J.~M., Jokipii, J.~R. \& Ellison, D.~C. \ 1996, \apj, 458,
641
\bibitem[\protect\citeauthoryear{Leroy et al.}{1982}]{leroy82}
Leroy, M. M., Winske, D., Goodrich, C. C., Wu, C. S., \&
Papadopoulos, K., \ 1982, \jgr, 87, 5081
\bibitem[\protect\citeauthoryear{Li et al.}{2009}]{lwsz09}
Li, G., Webb, G., Shalchi, A. \& Zank, G. P., \ 2009, 18th Annual
International Astrophysics Conference. AIP Conference Proceedings,
1183, 57
\bibitem[\protect\citeauthoryear{Malkov}{1997}]{malkov97}
    Malkov, M.~A., 1997, \apj, 485, 638
\bibitem[\protect\citeauthoryear{Malkov, Diamond \& V\"{o}lk}{2000}]{mdv00}
    Malkov, M.~A., Diamond P.~H. \& V\"{o}lk, H.~J., 2000, \apj Letters, 533, 171
\bibitem[\protect\citeauthoryear{Parker}{1961}]{parker61}
Parker, E.~N., \ 1961, Nucl. Energy, Vol.C2, 146
\bibitem[\protect\citeauthoryear{Ostrowski}{1991}]{Ostrowski91}
Ostrowski, M. , 1991, \mnras, 249, 551
\bibitem[\protect\citeauthoryear{ Shalchi, Li \& Zank }{2010}]{slz10}
Shalchi, A. ,  Li, G. \& Zank, G. P., \ 2010, Astrophysics and Space
 Science, 325, 99.
\bibitem[\protect\citeauthoryear{Vladimirov, Ellison \& Bykov}{2006}]{veb06}
    Vladimirov, A., Ellison, D.~C. \& Bykov, A., 2006, \apj, 652, 1246
\bibitem[\protect\citeauthoryear{Vladimirov, Ellison \& Bykov}{2008}]{veb08}
    Vladimirov, A., Ellison, D.~C. \& Bykov, A., 2008, \apj, 688, 1084
\bibitem[\protect\citeauthoryear{Zirakashvili \& Aharonian}{2010}]{za10}
 Zirakashvili, V.~N. \& Aharonian, F.~A., 2010, \apj, 708, 965
\end{thebibliography}

\newpage
\appendix
\section{Simulation box \& parameters }\label{appen-schematic}
\begin{figure}\begin{center}
  \includegraphics[width=3.0in, angle=0]{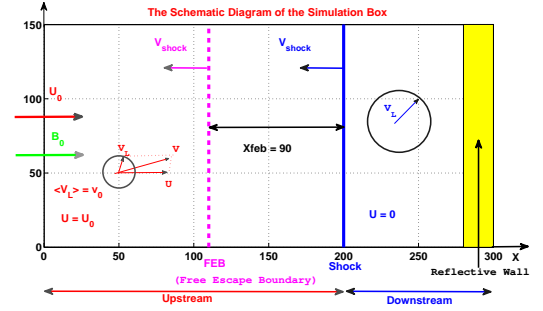}
  \end{center}
\caption{ Shock is produced by supersonic flow toward the stationary
reflective wall to the right. Continuous inflow of new particles
occurs at the left boundary. Inflow velocity is $U_{0}$, a
particle's total velocity is $V$, the local frame particle velocity
is $V_{L}$, and thermal velocity $<V_{L}>=v_{0}$. The two circles
represent one typical particle in the upstream region and one in the
downstream region, respectively. The vertical solid line represents
the shock front, the vertical dashed line represents the FEB, the
velocity of the shock is $V_{shock}$, size of the foreshock region
is $Xfeb=90$, the upstream flow velocity $U=U_{0}$, and the
downstream flow velocity is $U=0$. The magnetic field $B_{0}$ and
inflow velocity $U_{0}$ are both normal to the shock front
\citep[see][]{knerr96}.\label{schematicfig}}
\end{figure}

With respect to the validity and consistency of verifying the
previous dynamic Monte Carlo simulation method, the present
simulation program uses the same simulation box and identical
parameters as the previous dynamical Monte Carlo simulation
\citep{knerr96}. The schematic diagram of the simulation box is
shown in Figure \ref{schematicfig} and all the simulation parameters
are listed in Table \ref{parametertab}.
\begin{table*}[h]
  \centering
  \caption{\label{parametertab}The parameters of the simulated cases}
  \begin{tabular}{|c|c|c|c|}
  \hline
       Inflow velocity & $u_{0}$=0. 3& 403km/s  \\
\hline Thermal speed & $\upsilon_{0}$=0. 02 & 26. 9km/s \\
\hline Scattering time & $\tau$=0. 833 & 0. 13s  \\
\hline Box size & $X_{max}$=300 & $10R_{e} $ \\
\hline Total time & $T_{max}$=2400 & 6. 3minutes  \\
\hline Time step size & $dt$=1/30 & 0. 0053s  \\
\hline Number of zones & $nx$=600 & . . . \\
\hline Initial particles per cell & $n_{0}$=650 & . . .  \\
\hline FEB distance & $X_{feb}$=90 & $3R_{e}$ \\
\hline
\end{tabular}
 \tablefoot{Scaling used a  box size = 10$R_{e}$ (where $R_{e}$ represents the Earth's radius)
  and the box frame inflow velocity $u_{0}$=
 403km/s.
 This implies the following scale factors for distance,  velocity, and
 time:
  $X_{scale}$=10$R_{e}$/300, $v_{scale}$=403km $s^{-1}$/0.3,
  and $t_{scale}$=$x_{scale}$/$v_{scale}$. Here, the Mach number
  M=11.6. Dimensionless or normalized numbers are
used in the text to describe our simulations, except for
specifically highlighted examples \citep[see][]{knerr96}.}
  \end{table*}

\end{document}